\documentstyle[preprint,pre,aps]{revtex}
\begin{document}
\draft
\title{Scattering from supramacromolecular structures}
\author{Carlos I. Mendoza and Carlos M. Marques\\}
\address{L.D.F.C.- UMR 7506, 3 rue de l'Universit\'e, 67084 Strasbourg
Cedex, France}
\date{\today}
\maketitle
\begin{abstract}
We study theoretically the scattering imprint of a
number of branched supramacromolecular architectures, namely, polydisperse
stars and dendrimeric, hyperbranched structures. We show that polydispersity
and nature of branching highly influence the intermediate wavevector region
of the scattering structure factor, thus providing insight into the
morphology of different aggregates formed in polymer solutions.
\end{abstract}

% body of paper here
\noindent {\bf I. Introduction:}
\vskip 1pt

Scattering by light, x-rays or neutron radiation provides a fundamental tool
to investigate the shapes and statistical nature of large molecules in
solution\cite{guar}. For objects of fixed shape, like spherical, ellipsoidal
or cylindrical colloids, the scattering functions are known and can easily
be compared to experimental data, thus allowing for a determination of
shapes and relevant dimensions of the objects in a given experimental
sample. For objects of a relatively simple geometry, but with a fluctuating
nature, like polymer chains or semi-flexible rods the scattering spectra
contains not only information about the average shape of the mass
distribution but carries also a signature of the conformational disorder
determined by the nature of the thermodynamic fluctuations. For flexible
polymers, for instance, one can determine from the scattering data  whether
monomer-monomer excluded volume interactions are relevant or not in a
particular, given solvent.

Association of fixed-shape objects, like the aggregation of spherical
colloids that lead to fractal D.L.A.(diffusion-limited aggregation)
structures\cite{DLA}, brings  also some degree of disorder into the spatial
distribution of the scattering elements. Although the object as  a whole
does not fluctuate in time, the spatial distribution of the scattering
elements is statistically fixed by the aggregation process itself. D.L.A.
and other related processes have been shown to lead to self-similar
aggregates\cite{oh-sorensen}, where the frozen position  correlations $g({
r})$ between different elements at a distance $ r$ are described by a power
law $g({ r})\sim  r^{-m}$. It is well known that the scattering data from
these objects also carries the signature of the exponent $m$, thus providing
some insight on the type of the aggregation process ruling the solution
behavior.

When the aggregation process involves fluctuating objects,  the scattering
function carries information  both on the connectivity between different
scattering elements and on the statistics of the fluctuations. In this paper
we discuss the interplay between these two factors by studying a number of
branched polymer structures. Branched polymers are arguably the larger class
of systems of connected fluctuating elements, systems that also include
aggregates of soft gel beads, emulsion droplets and others. In many polymer
or polymer-like structures, control of the branching chemistry allows for a
careful choice  of the connectivity, like in dendrimers\cite{degennes,murat}
or in star-branched polymers\cite{mendes,al-muallem 1}, and to some extent in
dendrimeric (also called hyperbranched)
polymers\cite{knauss,sunder,muchtar}. Geometries with higher degree of
disorder are obtained by spontaneous aggregation in solutions of polymers
carrying sticking groups, and by random branching during polymerization
growth. This leads to a great diversity in the connections and to a
polydispersity in sizes of the constitutive elements. In order to clarify
the role of each of these factors in  shaping the scattering functions, we
explore a number of structures obtained by variations of star-like polymers:
stars with polydisperse arms, and dendrimeric polymers with different
degrees of polydispersity.

The paper is organized as follows. In section II we recall some basic
general features of the scattering amplitude for well known, fixed-shape and
fluctuating objects. Section III discusses the scattering function of
gaussian branched and hyperbranched structures, and also discusses
qualitatively the effects of excluded volume. Finally, in the conclusions we
discuss the different scattering signatures according to branching structure
and statistical nature of the fluctuations.

\vskip 3pt
%%%%%%%%%%%%%%%%%%%%%%%%%%%%%%%%%%%%%%%%%%%
\noindent {\bf II. Scattering from simple polymer architectures.}
\vskip 1pt The structure factor of an aggregate is given by
\begin{equation} S(\textbf{q})=\frac{1}{\cal
N}\biggl\langle\sum_{n,m=1}^{\cal N}\exp{\{i\textbf{q}\cdot(\textbf{
R}_{n}-\textbf{R}_{m})\}\biggr\rangle},
\label{g1}
\end{equation} where $\cal N$ is the number of scattering units in the
aggregate (monomers),
$\textbf{R}_{i}$ is the position of the $i$-th scattering unit, the ensemble
average is denoted by $\langle...\rangle$, and  $\textbf{q}$ is the momentum
transfer given by
$\textbf{q}=\textbf{q}_{s}-\textbf{q}_{i}$. Here $\textbf{q}_{i}$ and
$\textbf{q}_{s}$ are the wave vectors of the incident and scattered fields.
For elastic scattering
$\bracevert\textbf{q}_{s}\bracevert=\bracevert\textbf{q}_{i}\bracevert=2\pi/\lambda$,
where $\lambda$ is the wavelength of the incident wave. Hence,
$q\equiv\bracevert\textbf{q}\bracevert=(4\pi /\lambda)\sin{\theta /2}$, with
$\theta$ the scattering angle.

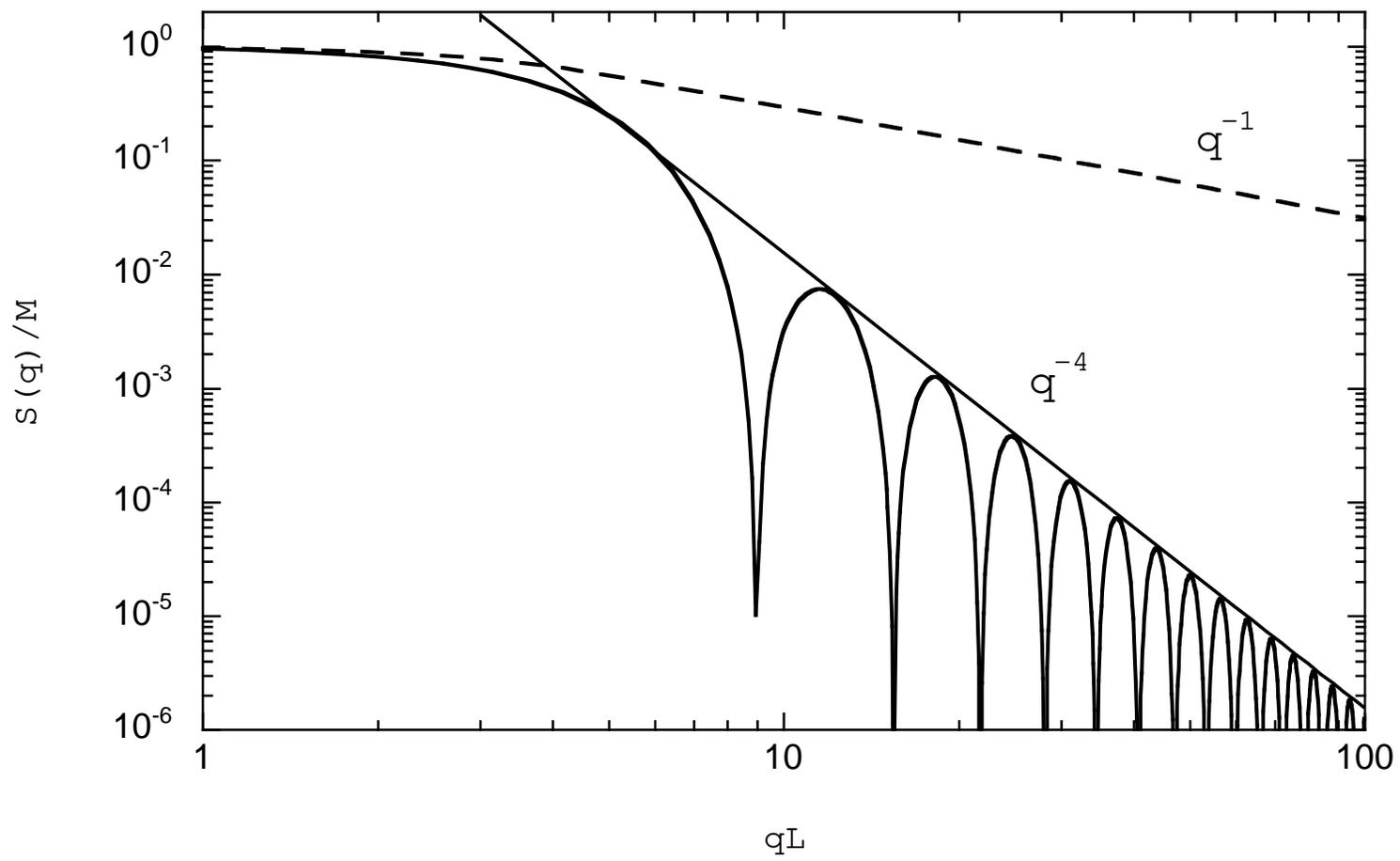
\begin{figure}
\caption{Sketch of the scattering intensity for a uniform rod of length
$L$ and mass $M$ (dashed line) and a uniform sphere of radius $R=L/2$  and
the same mass (solid line). A straight line with slope -4 is also shown.}
\label{rod-sphere}
\end{figure}

An illustration of the form of the scattering amplitude for two objects of
well defined shapes is shown in Fig.\ref{rod-sphere}. In this figure, the
spherically averaged scattering function of an infinitely thin uniform rod
of length $L$ is compared to that of an uniform sphere of radius $R=L/2$.
For simplicity both curves have been normalized to $1$ at the origin. The
curve corresponding to the rod shows a power law region with slope $-1$ for
large wavevector $q$, reflecting the unidimensional nature of the rod.
Indeed, for a mass density distribution of the form
$g(r)\sim r^{-(3-D)}$, where $D$ is the fractal dimension of the object, one
expects the scattering behavior $S(\textbf{q})\sim q^{-D}$. For rods,
$D=1$, and $S(\textbf{q})\sim q^{-1}$. This holds for spatial scales
$r\ll R_{g}$ or reciprocal lengths $q R_g \gg 1$, where $R_{g}$ is the
radius of gyration of the object
$R_{g}^2=(1/2{\cal N}^2)
\sum_{i,j}(R_{i}-R_{j})^2$. For rods, $R_{g}^2=L^2/12$. At small wavevectors
$qR_{g}\ll 1$, $S(\textbf{q})$ rolls off into a plateau (Guinier) region.
Deviations from the plateau value give a measure of the radius of gyration of
the object. In this region $S(\textbf{q})\sim 1-q^2 R_{g}^2/3$. The
structure factor of the sphere also shows a Guinier zone that rolls off from
the plateau at a smaller wavevector
$q$ indicating that the radius of gyration of this sphere is larger than the
radius of gyration of the rod. For larger values of $q$, the sphere
structure factor oscillates, the separation between the peak positions being
also a measure of the size of the sphere. The decaying envelope has a slope
$-4$, typical of objects with sharp interfaces.

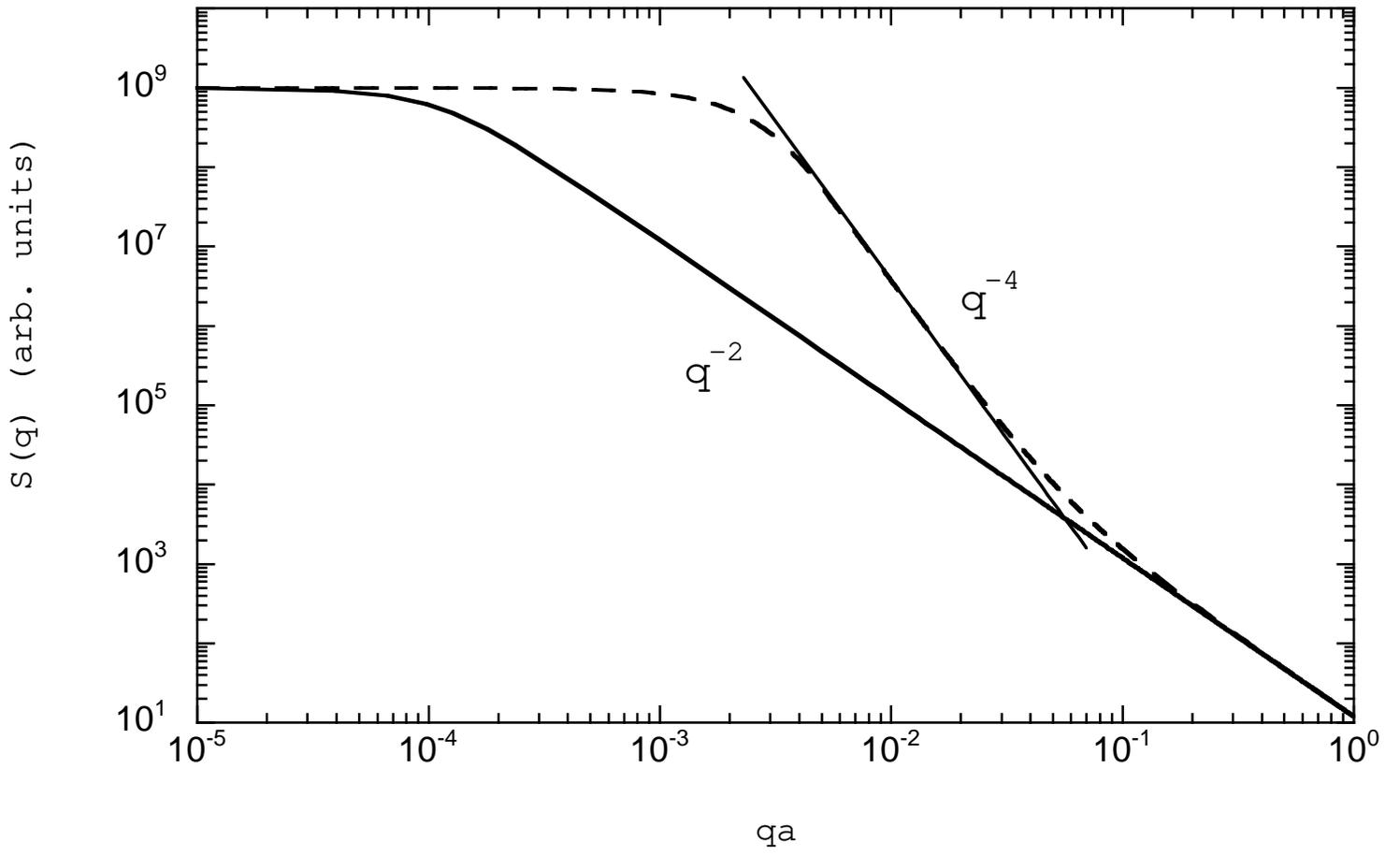
\begin{figure}
\caption{Scattering intensity for a linear Gaussian polymer of $P\times N$
monomers (solid line) and a Gaussian star of $P=10^{3}$  arms each one of
$N=10^{6}$ monomers (dashed line). A straight line with slope -4 is also
shown.}
\label{linear-star}
\end{figure}

As an example of the scattering function of fluctuating objects, we consider
the structure factor of a linear Gaussian polymer with $P\times N$ monomers
and of an uniform star of
$P$ arms of $N$ monomers each one [see Fig.\ref{linear-star} and also
Eq.(\ref{gstar})]. For the linear polymer, the large
$q$ behavior follows a power law with a slope
$-2$, a manifestation of the fractal dimension of a Gaussian chain,
$D=2$. The same behavior is seen at large wave vectors for the star
scattering function. However, the star shows a second power law region with
slope $-4$ before saturating at the plateau value. Notice that a steepest
curve is necessary to connect a plateau extending further in $q$ (the radius
of gyration of the star is smaller than the radius of gyration of the linear
polymer) to a coincidental high $q$ region in which both, the polymer and
the star, have the same local statistical structure. Quantitatively, the
value $-4$ is related to the average concentration profile. Indeed, it can
be shown\cite{star} that, when scattering from an inhomogeneous region of
average concentration
$\phi (r)$ dominates the spectrum, $S(\textbf{q})$ is calculated from
\begin{equation} S(\textbf{q})\sim\frac{1}{\cal N}\biggl[\int_0^{\infty} r^2
dr
\frac{\sin{qr}}{qr}
\phi (r)\biggr]^2.
\label{gstardaoud1}
\end{equation} For a Gaussian star polymer, the average concentration
profile is given by $\phi (r)=3P/(2\pi a^2 r)\  {\rm erfc}(r/2R_{g})$, where
$r$ is the distance from the center of the star,
$a$ is the monomer size, and ${\rm erfc}(x)$ is the complementary error
function\cite{abramowitz}. Equation (\ref{gstardaoud1}) thus leads to
\begin{equation} S(\textbf{q})\sim {P\over N q^{4}},\qquad 1\ll qR_{g}\ll
P^{1/2},
\label{gstardaoud2}
\end{equation} Note that this form crosses over correctly from the plateau
region
$S(q)\sim P\times N$ at
$q\sim R_g^{-1}$ to the high $q$-region $S(q)\sim q^{-2}$ at $q\sim P^{1/2}
R_g^{-1}
$.

Even in the absence of an exact form for the radial concentration profile, a
general argument can be made\cite{daoud-cotton} in order to extract the
value of the intermediate slope. For a star with $P$ arms, the average
concentration inside an infinitesimal spherical shell of volume $4\pi r^2
dr$ and radius $r$, centered on the star, is given by
\begin{equation}
\phi (r)=PdN(r)/(4\pi r^2 dr),
\label{phi1}
\end{equation} with $dN(r)$ is the average number of monomers per arm in the
shell. For a Gaussian arm $N(r)\simeq 3r^2 /a^2$ and one easily recovers the
exact result for the concentration in the scaling regime $r\ll R_{g}$. The
Daoud and Cotton argument (\ref{phi1}), combined with
equation(\ref{gstardaoud1})  also allows to compute the intermediate regime
of the scattering factor of a star in a good solvent, where excluded volume
interactions are important. In this case, the star can be described as a
semi-dilute solution, with a local, position dependent screening length $\xi
(r)$. Pictorially, this is represented by arms made of a succession of blobs
of increasing size $\xi (r)$. The arms are stretched and within a distance
$r$ from the center one finds
$ N(r)  \simeq (r/a)^{5/3}P^{-1/3}$ monomers, the size of the star in a good
solvent being $R_{\rm star}\simeq N^{3/5} P^{1/5}$. It follows that the
average concentration varies radially as
$\phi (r)\sim P^{2/3} r^{-4/3}$ and the intermediate regime is described by
\begin{equation} S(\textbf{q})\sim {P^{1/3}\over N q^{10/3}},\qquad 1\ll q
R_{\rm star}\ll P^{2/5}.
\label{gstardaoud3}
\end{equation}
    For higher wavevectors, $q R_{\rm star}\gg P^{2/5}$, the function
$S(\textbf{q})$ crosses over to a slope
$-5/3$ indicative of the local excluded volume statistics of the
arms\cite{plate}.

\vskip 3pt
\pagebreak
%%%%%%%%%%%%%%%%%%%%%%%%%%%%%%%%%%%%%%%%%%%
\noindent {\bf III. Branched structures:}
\vskip 1pt The objective of this section is to show how the internal branched
structure of the aggregate modifies the form of the structure factor. In
order to do this, we consider two different types of aggregates, (a)
polydisperse stars, and (b) dendrimeric structures (see
Fig.\ref{structures}).

The method
of calculation for the structure factor of aggregates (a) and (b) starts by
considering an arbitrary aggregate whose scattering function is  known. If a
new arm is added to the structure, the scattering function that accounts for
the new arm can be calculated in terms of the known scattering function plus
corrections due to the correlations between the monomers of the new arm with themselves
and with all the monomers of the previous aggregate. A more detailed discussion of
the procedure is  given in Appendix A where the structure factor of a
dendrimer of two generations is computed in terms of the structure factor of
a monodisperse star.

{\em III.(a) Polydisperse stars}

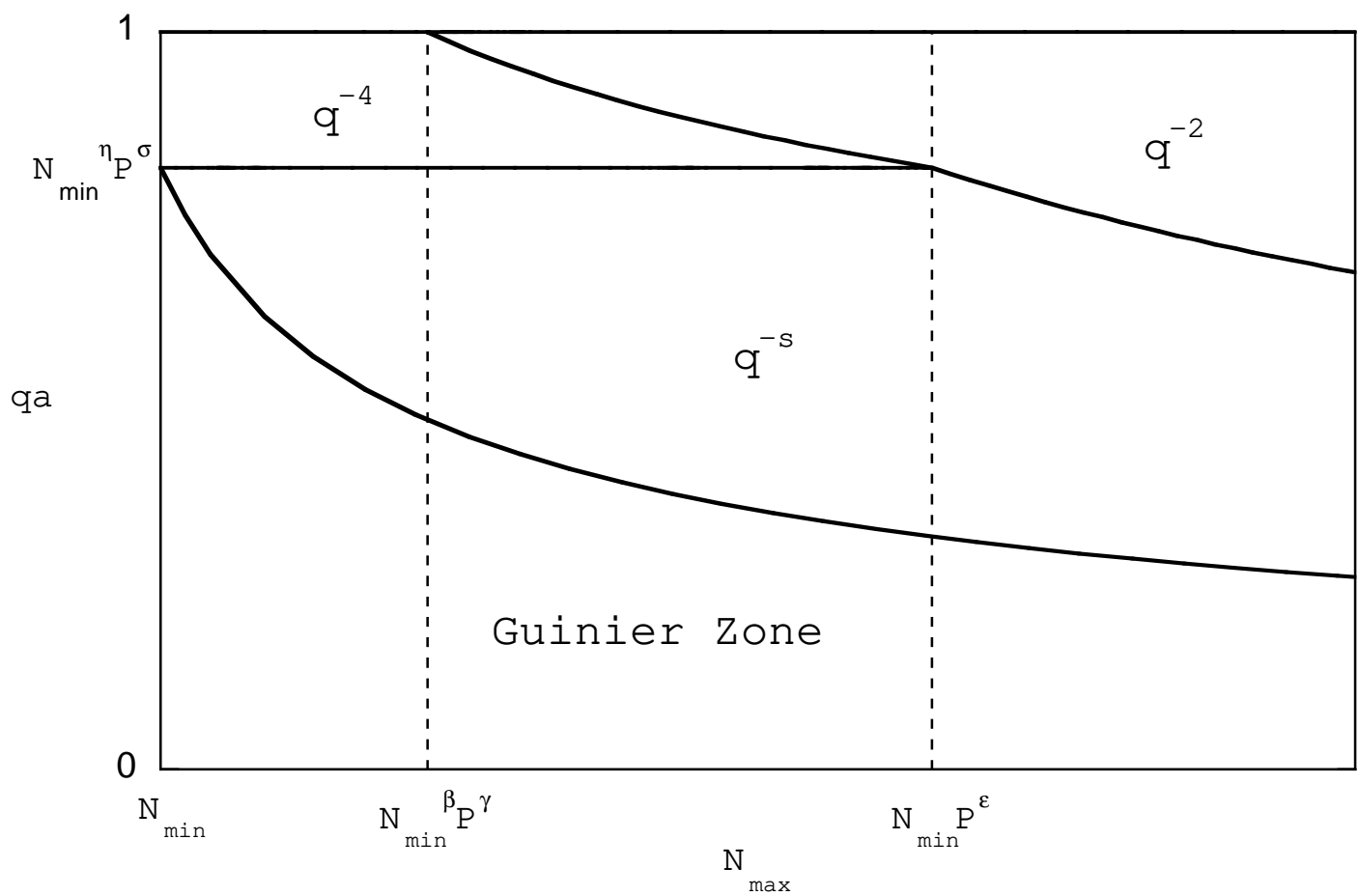
\begin{figure}
\caption{Diagram showing the different scaling regimes of the scattering
function for: polydisperse stars in
$\theta$ solvent [$\beta =-1/(m-1)$, $\gamma =m/(m-1)$, $\epsilon =\gamma$,
$\eta =-1/2$, $\sigma =0$, and $s=2(2-1/m)$], polydisperse stars in athermal
solvent [$\beta =-4/(6m-5)$, $\gamma =1/3(6m-1)/(6m-5)$, $\epsilon
=2\gamma$, $\eta =-3/5$, $\sigma =-1/5$, and $s=10/3(1-2/(6m-1))$],
dendrimeric structures in
$\theta$ solvent [$\beta =1/(2m+1)$, $\gamma =2m/(2m+1)$, $\epsilon
=\gamma$, $\eta =-1/2$,
$\sigma =0$, and
$s=2(2+1/m)$], and dendrimeric structures in athermal solvent [$\beta
=1/(2m+1)$, $\gamma =(1/3)2m/(2m+1)$, $\epsilon =2\gamma$, $\eta =-3/5$,
$\sigma=-1/5$, and
$s=10/3(1+2/(6m+1))$].}
\label{regimes}
\end{figure}

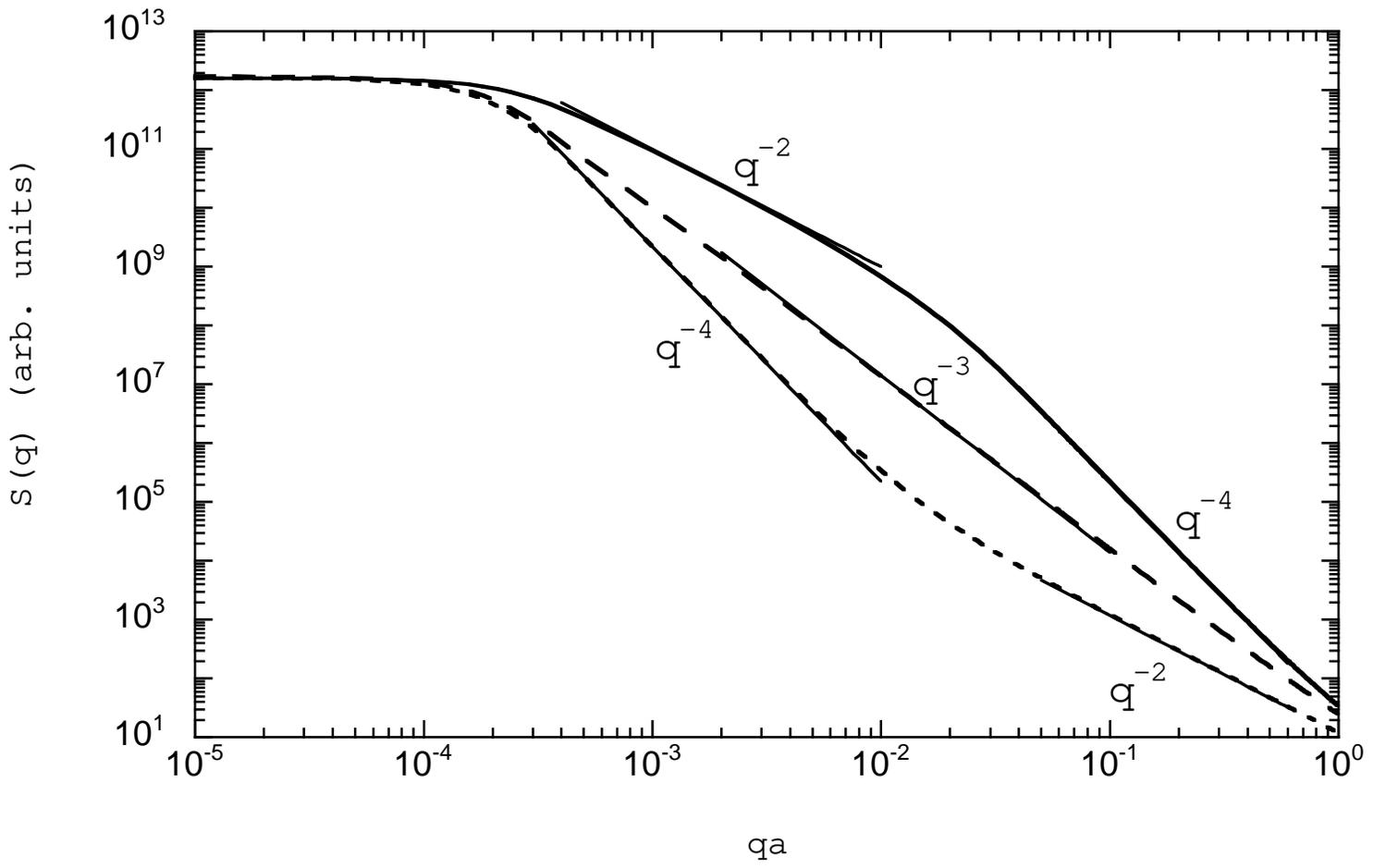
\begin{figure}
\caption{A comparison of the scattering intensity as a function of the
scattering vector in a log-log scale for polydisperse stars of the same mass
in $\theta$ solvent. The parameters for the polydispersed  stars are
$G=100$, $P_{i}=10^{4}$, and $N_{i}\sim 10^{8}\times i^{-2m}$ with $m=1$
(solid line), $m=2$ (dashed line) and $m\rightarrow \infty$ (dotted line).
This last case corresponds to a uniform star of $P_{G}=10^4$ arms. Straight
lines with slopes $-4$, $-3$ and $-2$ are also shown.}
\label{gstars}
\end{figure}

Consider a star-branched polymer made of $P_1$ arms of $N_1$ monomers, $P_2$
arms of
$N_2$ monomers and so on (see Fig.\ref{structures}a). The structure factor
of this aggregate is
\begin{eqnarray} S({\bf q})
&=&\frac{1}{\sum_{i=1}^{G}P_{i}N_{i}}\sum_{p,q=1}^{G}
P_{p}P_{q}\frac{N_{p}N_{q}}{x_{p}x_{q}}(\exp(-x_{p})-1)(\exp(-x_{q})-1)\nonumber\\
&&+\frac{1}{\sum_{i=1}^{G}P_{i}N_{i}}\sum_{q=1}^{G}
P_{q}\frac{N_{q}^2}{x_{q}^{2}}
\biggl[ 2 (\exp(-x_{q})-1+x_{q})-(\exp(-x_{q})-1)^2\biggr].
\end{eqnarray} Here, $G$ is the number of different lengths,
$x_{q}=q^2a^2N_{q}/6$.  Note that this expression can be decomposed, as for
the monodisperse star, in a contribution from the average concentration of
the star plus contributions from the fluctuations. However, as we will see
below, the general shape of the scattering curve now exhibits a richer
behavior. The low wavevector  and the high wavevector regions still present
the  usual Guinier roll-off from a plateau and a `` -2 slope'',
respectively. The signature of the star polydispersity is carried by the
shape of the intermediate scattering region.  Consider the case of arm-size
polydispersity. Here, we take the star to be made of
$G$ different stars, each of them with equal arm number $P$ but an arm-size
distribution $N_i$. For the form $N_i=N_{max} i^{- 2 m}$, where $m \ge 0$,
$i$ runs from $1$ to $G$ and $N_{max}$ is the number of monomers of the
largest arms, the radius of  gyration of each arm is
$R_i = {\rm Cte}\ i^{-m} a$, and the average concentration can be written
as\cite{highp}
\begin{equation}
\phi (r)= \sum_i \phi_i (r)={3P\over 2\pi a^2 r} \sum_i \  {\rm
erfc}(r/2R_{i}) \simeq {3P\over 2\pi a^3 } \frac{{(2 \ {\rm
Cte})^{{\frac{1}{m}}}}\Gamma({\frac{1 + m}{2\,m}})}
     {{\sqrt{\pi }}\,} \left({a\over r}\right)^{m+1\over m},
\label{phipoly}
\end{equation} where $\Gamma$ is the gamma function $ \Gamma (a) =
\int_0^\infty\ dx  x^{a-1} \exp\{ - x\}$. If the exponent $m$ is very large,
only the $P$ largest arms  contribute to the concentration profile which is
very similar to a monodisperse star. As the distribution width increases,
the smaller arms start to significantly  contribute to the profile. In this
case, several different regimes can be identified, as shown in
Fig.(\ref{regimes}). By inserting the continuous limit of Eq.(\ref{phipoly})
in Eq.(\ref{gstardaoud1}) we find an intermediate regime that scales as
$S(q) \sim q^{-2(2-1/m)}$, for $m \ge 1/2$. This means that for the
polydispersity considered here, the exponent varies from $0$ when $m=1/2$ to
$-4$ when $m \rightarrow \infty$. Interestingly, a second intermediate
regime develops due to the finite distribution of the arm polydispersity.
When the size of arms has a finite small cutoff at $N_{min}=N_G$, the
polydisperse star consist of two parts: the outer polydisperse shell and a
central star-like core. This second intermediate regime therefore appears at
length scales smaller than the length of the smaller arms and follows the
usual $q^{-4}$ scaling. At low wavevectors $q$ the spectrum shows a Guinier
plateau. The transition point between the Guinier zone and the first
intermediate region occurs at
$(qR_g)^{2}\sim 1$ as can be verified by inserting the exact form of
Eq.(\ref{phipoly}) in Eq.(\ref{gstardaoud1}) and exploring the small
$q$ limit of the resulting expression.  The radius of gyration of the
polydisperse star is
$R_g^2= [\int_0^\infty dr\ r^4 \phi (r)]/[\int_0^\infty dr\ r^2
\phi (r)] = 3 R_1^2 [\sum_{i} i^{-4 m}]/[\sum_{i} i^{-2 m}]$ where $R_1 \sim
a N_{1}^{1/2}$ is the gyration radius of the largest arm. For large values
of $m$ one recovers the limit of the  monodisperse star $R_g^2= 3 R_1^2$.
Written in terms of the monomer size $a$, the transition point between the
Guinier zone and the first intermediate regime occurs at $(qa)^{2}\sim
1/N_{max}$, where $N_{max}=N_{1}$ is the number of monomers of the largest
arms. The transition point between the two intermediate regions occurs at
$(qa)^{2}\sim 1/N_{min}$, where
$N_{min}=N_{G}$ is the number of monomers of the shortest arms. Finally, the
transition between the second intermediate region and the region where
fluctuations dominate occurs at
$(qa)^{2}\sim P/N_{max}(N_{max}/N_{min})^{1/m}$. One can see that the region
where fluctuations dominates appears only if $N_{max}>P^{m/(m - 1)}
N_{min}^{-1/(m - 1)}$. If the exponent $m$ is larger than one and the number
of monomers
$N_{max}$ is very large,
$N_{max}\gg N_{min}P^{m/(m-1)}$, the region with slope $q^{-4}$ disappears
and one crosses over directly to the region dominated by fluctuations with
slope
$q^{-2}$. The transition point then occurs at $(qa)^2 \sim
P^{m/(m-1)}/N_{max}$ (see Fig.\ref{regimes}). For values of $m$ between $1$
and $1/2$, the scattering function always crosses from the regime $S(q) \sim
q^{-2(2-1/m)}$ to a regime $S(q)
\sim q^{-4}$ independently of the number of monomers
$N_{max}$. Examples of some of these cases are presented in Fig.
\ref{gstars} (see note \cite{highp}). In the limit $m=1/2$ the scattering
function does not exhibit a developed scaling but rolls over gently from the
Guinier plateau to the large wavevector limit of slope $-2$.

For a polydisperse star in a good solvent a Daoud-Cotton argument can also
be applied  to determine the scattering regimes. Let $p(r)$ be the number of
arms at  distance $r$ from the center. Then, the local correlation length is
set by $\xi(r)= r  p(r)^{-1/2}$ and the concentration is $\phi(r) =
p(r)^{2/3} r^{-4/3}$. In the  infinitesimal shell of volume $dV = 4
\pi r^2 dr$, there are then $dn = p(r)^{- 1/3} r^{2/3} dr$ monomers  in one
arm. By knowing the polydispersity of the chains $p(n)$ one can compute the
function
$p(r)$ and extract the concentration. If we choose, as for the precedent
Gaussian example,
$N_i = N_{max} i^{- 2m}$, and a constant arm number of each length,
$P_i=P$, we get in the  continuous limit $p(n) = P (n/N_{max})^{-1/2m}$,
leading to $p(r) \sim r^{- 5/(6 m -1 )}$.  Correspondingly, the concentration
scales as $\phi(r)\sim r^{-\alpha}$ with the exponent
$\alpha={2\over 3}(2 + {5\over 6m-1})$ and the scattering from that
intermediate region scales as $S(q)\sim  q^{-s} $, with exponent
$ s= {10\over 3}(1 - {2\over 6m-1})$. In the limit where $m$ is very  large
one recovers the usual intermediate slope $s= 10/3$. As $m$ decreases we
reduce the  intermediate slope, as for the Gaussian case.  The transition
point between the Guinier zone and this region occurs at $(qa)^{5/3}\sim
1/(N_{max}P^{1/3})$. As in the Gaussian case, we find a second intermediate
region that scales as $S(q)\sim q^{-10/3}$. The transition point between the
two intermediate regions occurs at
$(qa)^{5/3}\sim 1/(N_{min}P^{1/3})$. Finally, the transition between the
second intermediate region and the fluctuating regime occurs at
$(qa)^{5/3}\sim P^{1/3}/N_{max} (N_{max}/N_{min})^{4/(6m-1)}$. These results
are shown in Fig.\ref{regimes}. One can see that the region where
fluctuations dominates appears only if $N_{max}>P^{1/3(6m-1)/(6m-5)}
N_{min}^{-4/(6m-5)}$. If the exponent $m\ge 5/6$ and the number of monomers
$N_{max}$ is very large, $N_{max}\gg N_{min}P^{2/3 (6m-1)/(6m-5)}$, the
region with slope $q^{-10/3}$ disappears and one crosses over directly to
the region dominated by fluctuations with slope
$q^{-5/3}$. The transition point then occurs at $(qa)^{5/3} \sim
P^{(2m+1)/(6m-5)}/N_{max}$. Again, by choosing the value of $m$ between
$\infty$ and
$5/6$, it is  possible to obtain scattering functions with intermediate
slopes ranging from the monodisperse star value of
$-10/3$ to $-5/3$ slope. For values of $m$ between $5/6$ and $1/2$, the
scattering function always crosses from the regime $S(q) \sim
q^{-10/3(1-2/(6m-1))}$ to a regime $S(q) \sim q^{-10/3}$ independently of
the number of monomers $N_{max}$.

{\em III.(b)Dendrimeric Structures}

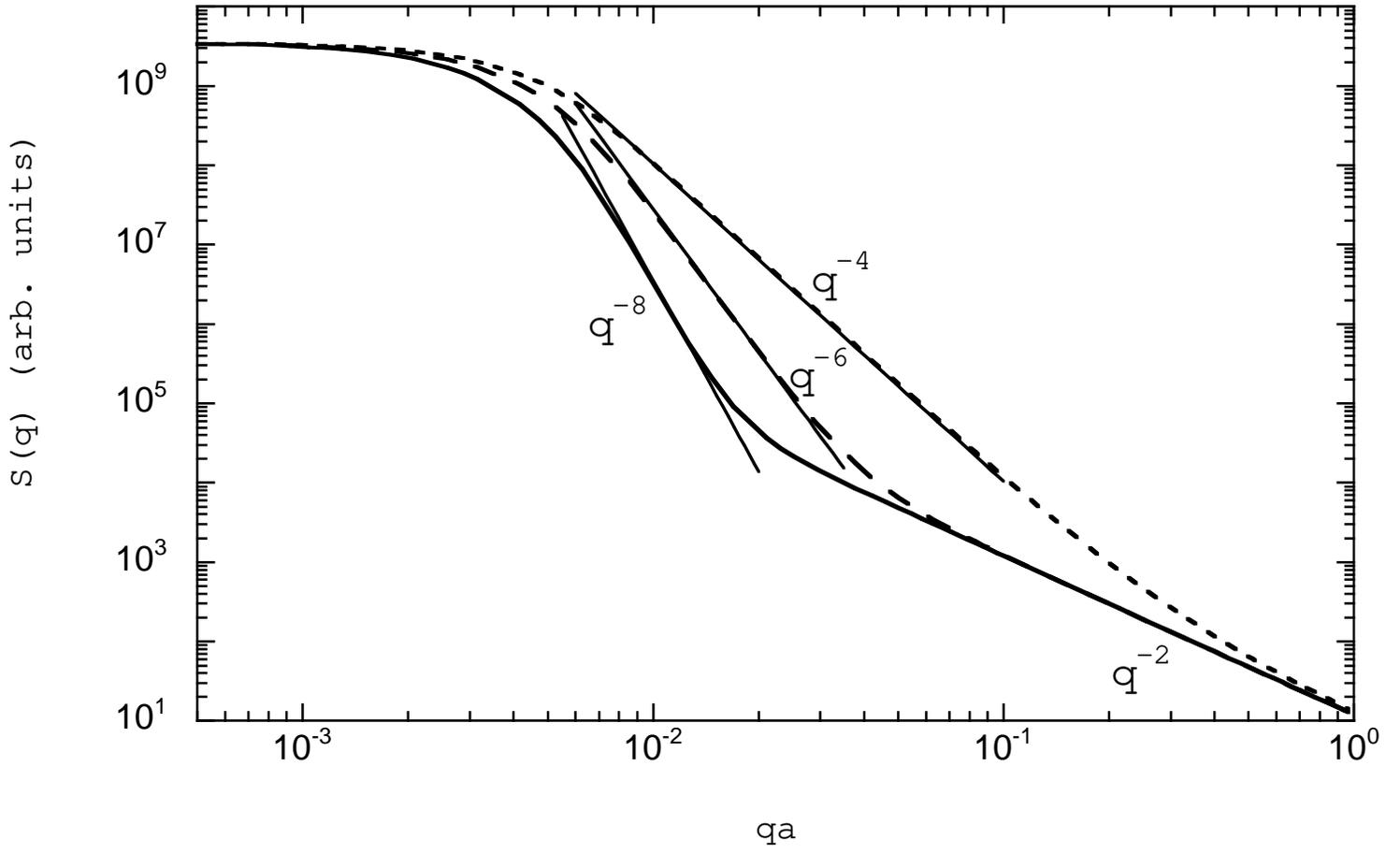
\begin{figure}
\caption{A comparison of the scattering intensity as a function of the
scattering vector in a log-log scale for dendrimers with the same mass. The
parameters used were $G=10$,
$P=20$, and
${\cal N}_{i}\sim 10^{3}\times i^{-2m}$ with $m=1/2$ (solid line), $m=1$
(dashed line) and $m\rightarrow \infty$ (dotted line). This last case
corresponds to a uniform star of $P_{G}=20\times 2^9$ arms. Straight lines
with slopes
$-8$, $-6$ and $-4$ are also shown.}
\label{gdendrimers-scaling}
\end{figure}

These structures are formed by starting from a uniform star of $P_{1}=P$
arms of $N_{1}$ monomers each one, and branching each arm twice so that in
the second generation there are
$P_{2}=2P_{1}$ arms of length $N_{2}$. We then branch each of the newest
arms twice so that in the third generation there are
$P_{3}=2P_{2}=2^{2}P_{1}$ arms of length $N_{3}$. We repeat this process up
to any desired number of generations (see Fig.\ref{structures}b). This means
that the number of arms in each generation is $P_{i}=2^{i-1} P$. As in the
case of the polydisperse stars, for a large number of arms and generations,
asymptotic shapes can be reached for particular types of polydispersity
distributions. Let us consider the case of a dendrimer of $G$ generations
with arm-size polydispersity of the form ${\cal N}_{i}=2^{2m(i-1)} {\cal
N}_{min}$, where ${\cal N}_{i}=\sum_{j=1}^{i}{N_{j}}$ is the sum of monomers
per arm from generation
$1$ up to generation $i$. By choosing $m\ge 1/2$, we assure that the first
generation always has the smallest number of monomers per arm. By using
arguments similar to the ones for the polydisperse star, we find for the
Gaussian dendrimer, an average concentration $\phi(r) \sim r^{-(1-1/m)}$,
that gives rise to an intermediate scaling regime
$S(q)\sim q^{-2(2+1/m)}$\cite{concentration}. Again, if
$m$ is very large, only the last generation contributes to the scattering,
which is similar to the monodisperse star with
$P_{G}$ arms. As $m$ decreases, the first generations start to contribute
significantly, modifying the slope of the scattering curves that reaches the
steepest value of $-8$ when
$m=1/2$. The transition point between the Guinier zone and this region
occurs at
$(qa)^{2}\sim 1/{\cal N}_{max}$, where ${\cal N}_{max}={\cal N}_{G}$. Also,
there is a second intermediate region which scales as
$S(q)\sim q^{-4}$. The transition point between these two intermediate
regions occurs at
$(qa)^{2}\sim 1/{\cal N}_{min}$, where ${\cal N}_{min}={\cal N}_{1}$ (see
Fig.\ref{regimes}). Finally, the transition between the second intermediate
region and the fluctuating regime occurs at $(qa)^{2}\sim P/{\cal
N}_{max}({\cal N}_{max}/{\cal N}_{min})^{-1/2m}$. The region where
fluctuations dominates appears only if ${\cal N}_{max}>P^{2m/(2m+1)} {\cal
N}_{min}^{1/(2m+1)}$. The regime with $S(q)\sim q^{-4}$ disappears when
${\cal N}_{max}
\sim {\cal N}_{min} P^{2m/(2m+1)}$. In this case the transition point
between the region with $S(q)\sim q^{-2(2+1/m)}$ and the fluctuating regime
occurs at $(qa)^2 \sim P^{m/(m+1)}/{\cal N}_{max}({\cal N}_{max}/{\cal
N}_{min})^{1/2(m+1)}$. Note that for large ${\cal N}_{max}$ and by tuning
$m$ between
$\infty$ and $1/2$, it is possible to obtain scattering functions with
intermediate slopes ranging from the monodisperse star value of $-4$ to $-8$
slope. In Fig.\ref{gdendrimers-scaling} we show examples of some of these
cases.

Applying a Daoud-Cotton argument to the case of dendrimers in good solvent
we determine the corresponding scaling regimes. In this case, $p(r) \sim r^{
5/(6 m +1)}$.  Correspondingly, the concentration scales as $\phi(r)\sim
r^{-\alpha}$, with $\alpha ={2\over 3}(2 - {5\over 6m+1})$ and the
scattering from that intermediate region scales as $S(q)\sim q^{-s}$, with
$s={10\over 3} (1+{2\over 6m+1})$. In the limit where $m$ is very large one
recovers the usual intermediate slope $s= 10/3$. As
$m$ decreases we reduce the intermediate slope, as for the Gaussian case.
The transition point between the Guinier zone and this region occurs at
$(qa)^{5/3}\sim 1/(P^{1/3}{\cal N}_{max})({\cal N}_{max}/{\cal
N}_{min})^{-1/6m}$. As in the Gaussian case, we find another intermediate
region that scales as
$S(q)\sim q^{-10/3}$. The transition point between the two intermediate
regions occurs at
$(qa)^{5/3}\sim 1/(P^{1/3}{\cal N}_{min})$. Finally, the transition between
the second intermediate region and the fluctuating regime occurs at
$(qa)^{5/3}\sim P^{1/3}/{\cal N}_{max} ({\cal N}_{max}/{\cal
N}_{min})^{-1/2m}$. These results are shown in Fig.\ref{regimes}. The region
where fluctuations dominates appears only if ${\cal N}_{max}>{\cal
N}_{min}^{1/(2m+1)} P^{2m/(3(2m+1))}$. If
${\cal N}_{max}$ is very large, ${\cal N}_{max}\gg {\cal
N}_{min}P^{4m/(3(2m+1))}$, the region with slope
$q^{-10/3}$ disappears and one crosses over directly to the region dominated
by fluctuations with slope
$q^{-5/3}$. The transition point then occurs at $(qa)^{5/3} \sim
P^{(2m-1)/(6m+5)}/{\cal N}_{max}({\cal N}_{max}/{\cal
N}_{min})^{(2m-1)/(2m(6m+5))}$. Again, by choosing the value of $m$ between
$\infty$ and
$1/2$, it is  possible to obtain scattering functions with intermediate
slopes ranging from the monodisperse star value of $-10/3$ to $-5$ slope.

The results for the scaling regimes and the transitions between these
regimes for both, polydisperse stars and hyperbranched structures are
summarized in Appendix B.

In Fig.\ref{gdendrimer-generations} we plot the structure factor for
Gaussian dendrimers with arms of equal size for all the generations. In this
case a non-scaling regime is present between the Guinier zone and the high
$q$ region.  This region shows a hump that reflects the fact that as we
increase the number of generations, the outer core of the aggregate becomes
very dense thus dominating the structure of the spectrum which resembles the
one for a spherical shell. We see in Fig.\ref{gdendrimer-generations} that
the size of the hump increases as we increase the number of generations in
qualitative agreement with the experimental results of Ref.\cite{prosa}.
Note that while growth of dendrimers to a high number of generations is
usually hindered by steric reasons, a polydisperse dendrimer can grow
indefinitely if the polydispersity is correctly tuned.

From the scattering curves of branched structures in which there is a $q$
region where fluctuations dominate, we note that there is always an
intermediate $q$ region with a steeper slope than the corresponding one for
a linear polymer. This can be understood by using a simple graphical
argument. Consider a linear Gaussian polymer of mass
$N$. Its structure factor consist of a power law region with slope $-2$ and
a Guinier zone for low wave vector $q$. Now, let us consider any branched
Gaussian polymer. Its structure factor coincides, in slope and absolute
value, with the one for the linear Gaussian polymer at short wave lengths.
This reflects the fact that the internal structure of the polymer is the
same in both cases. If the branched polymer has the same mass than the
linear one, both spectra must coincide in the plateau Guinier zone. However,
since the radius of gyration of any branched polymer is always smaller than
the corresponding one for the linear polymer, the Guinier zone must extend
to a larger wave vector value, as shown for example in
Fig.\ref{linear-star}.  Therefore, the only possible way of crossing from
one region to the other is by an intermediate region with an average slope
larger than the slope at short-wave lengths. This argument is also valid in
good solvent conditions where the Gaussian model is not valid and
monomer-monomer interactions play an important role.

\vskip 3pt
%%%%%%%%%%%%%%%%%%%%%%%%%%%%%%%%%%%%%%%%%%%
\noindent {\bf IV. Conclusions:}
\vskip 1pt In this paper we have shown how different branched polymers give
rise to different structure factors. This information can be used to probe
the morphology of supramacromolecular aggregates. We have shown how the
slope in the intermediate $q$ region can be tailored according to the
polydispersity in the length of the constitutive linear chains of the
branched aggregates. In particular, for polydisperse stars we found scaling
regimes with slopes ranging from
$-2$ to $-4$ in $\theta$ solvent conditions and between $-5/3$ and $-10/3$
for athermal solvent. In the case of dendrimeric structures, scaling regimes
ranging from $-4$ to $-8$ in $\theta$ solvent and between $-10/3$ and
$-5$ for athermal conditions, although richer behavior was obtained for
specific choices of the polydispersity parameters. We have shown using
simple arguments that whenever there is a region where fluctuations dominate
the scattering response, then the structure factor of branched structures
always present an intermediate $q$ regime with at least a small region where
the slope in a log-log plot is larger than the corresponding slope of the
linear polymer. This means that these aggregates are not strictly
self-similar over the entire range of length scales $a<l<R_{g}$. Results
presented in this paper can be qualitatively used as a guiding tool for
exploring the branching morphology of aggregates according to the type of
regimes presented in the scattering intensity curves. They also provide
qualitative information from the analysis of the values of the slopes of the
intermediate $q$ regimes.

\vskip 3pt
%%%%%%%%%%%%%%%%%%%%%%%%%%%%%%%%%%%%%%%%%%%
\noindent {\bf Appendix A: Method of Calculation}
\vskip 1pt In this appendix we outline the procedure to obtain the structure
factor for the branched structures described in this paper by considering a
specific example. Suppose a branched polymer that grows following a given
rule like the one shown in Fig.
\ref{structures}b. In this figure, we show a polymer that grows from a star
of $P_{1}$ arms each one made of
$N_{1}$ monomers. Each arm is then branched in two other arms made of
$N_{2}$ monomers. Then, each arm of the newest generation is branched again
in two arms and so on. The structure factor of the structure made of $G$
generations is related to the structure factor of the structure made of
$G-1$ generations by the equation
\begin{equation}
S_{G}(\textbf{q})=\frac{\sum_{i=1}^{G-1}P_{i}N_{i}}{\sum_{i=1}^{G}P_{i}N_{i}}S_{
G-1}(\textbf{q}) +S_{corr}(\textbf{q}),
\end{equation} where $P_{i}$ and $N_{i}$ are the number of arms and monomers
in the $i$-th generation, respectively, $S_{i}(\textbf{q})$ is the structure
factor of the structure made of $i$ generations and the $corrections$,
$S_{corr}(\textbf{q})$, are due to the correlations between the arms grown
in the
$G$-th generation with themselves and the rest of the arms. Assuming that
$S_{G-1}(\textbf{q})$ is known, the problem consist in calculate these
corrections. Then, applying the procedure recursively, the structure factor
of any structure made of an arbitrary number of generations can be
calculated.

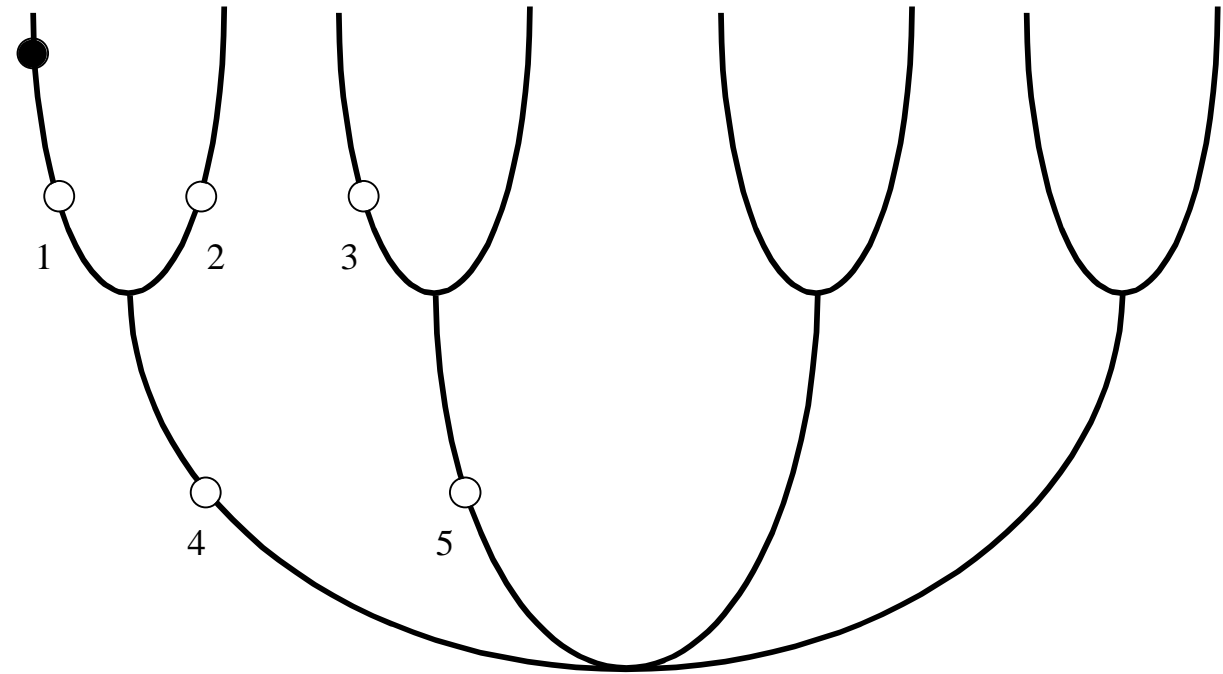
\begin{figure}
\caption{Representation of a dendrimer with $P=4$ arms, and $G=2$
generations. The solid circle represents the location of monomer $m$ and the
open circles represent the locations of monomers $n$ representatives of the
different families.}
\label{appendix}
\end{figure}

Now we proceed with the calculation of the $corrections$. In order to show
the general idea, we are going to treat the simple case of a dendrimer with
$G=2$ generations (see Fig.\ref{appendix}). Any monomer $m$ that belongs to
the last generation
$(G=2)$ interacts with all other monomers of the structure. These
interactions can be classified according to the relative location, in the
structure, of the second monomer $n$  with respect to to monomer $m$. This
is shown in Fig.\ref{appendix} where we classify the position of the second
monomer in 5 different families. The monomers $n$ of family 1 belongs to the
same generation and to the same arm of monomer $m$. Family 2 comprises all
the monomers $n$ of the same generation of $m$ but that belong to a
different arm whenever this arm has a common origin with the arm where
$m$ is located. Family 3 consists of monomers $n$ of the same generation of
$m$ and whose respective arms originate in different arms of the previous
generation. For family 4 one monomer, say $m$, belongs to generation $G$ and
the other $(n)$ belongs to generation $G-1$, and the arm where $m$ is located
originates in the arm where $n$ is located. Finally, family 5 consists of
monomers $n$ of different generation than that of monomer $m$, and the
monomer $m$ can not be reached by going through the arm where $n$ is
located. All the monomers that  participate in the corrections belong to one
of these families. Contributions to the structure  factor from a given family
$f$ are calculated from the expression
\begin{equation} S_{corr}^{f}(\textbf{q})=\frac{1}{N_{T}}\sum_{m,n}\int
d\textbf{R}_{m}d\textbf{R}_{n}G_{m,n}^{f}(\textbf{R}_{n},\textbf{R}_{m})
\exp{\{i\textbf{q}\cdot(\textbf{R}_{n}-\textbf{R}_{m})\}}
\label{gappendix}
\end{equation} where $N_{T}$ is the total number of monomers, the sum runs
over monomer $m$ and $n$ with at least one of them  belonging to the latest
generation ($G=2$), and $G_{m,n}^{f}(\textbf{R}_{n},\textbf{R}_{m})$  is the
Green function of the
$f$ family given by
\begin{equation}
G_{m,n}^{1}(\textbf{R}_{n},\textbf{R}_{m})=G(\textbf{R}_{n},\textbf{R}_{m};m-n),
\end{equation}

\begin{equation}
G_{m,n}^{2}(\textbf{R}_{n},\textbf{R}_{m})=G(\textbf{R}_{n},\textbf{R}_{m};m+n),
\end{equation}

\begin{equation}
G_{m,n}^{3}(\textbf{R}_{n},\textbf{R}_{m})=G(\textbf{R}_{n},\textbf{R}_{m};m+n+2N_{1}),
\end{equation}

\begin{equation}
G_{m,n}^{4}(\textbf{R}_{n},\textbf{R}_{m})=G(\textbf{R}_{n},\textbf{R}_{m};m-n+N_{1}),
\end{equation}

\begin{equation}
G_{m,n}^{5}(\textbf{R}_{n},\textbf{R}_{m})=G(\textbf{R}_{n},\textbf{R}_{m};m+n+N_{1}),
\end{equation} where
\begin{equation}
G(\textbf{R}_{n},\textbf{R}_{m};\alpha)=\biggl(\frac{3}{2\pi a^2\bracevert
\alpha\bracevert}\biggr)^{(3/2)}\exp{\biggl\{-\frac{3(\textbf{R}_{m}-\textbf{R}_{n})^2}{2
a^2\bracevert
\alpha\bracevert}\biggr\}}.
\end{equation}

Substituting these expressions in Eq.(\ref{gappendix}) and taking the
continuos limit by transforming the sums into integrals we find the structure
factor for the dendrimer of two generations
\begin{eqnarray} S_{2}(\textbf{q})
&=&\frac{N_{1}}{N_{1}+2N_{2}}S_{1}(\textbf{q}) \nonumber\\
&&+\frac{2N_{2}}{N_{1}+2N_{2}}
\biggl[\frac{N_{2}}{x_{2}^2}(\exp{\{-x_{2}\}}-1)^2
+\frac{2N_{2}}{x_{2}^2}(\exp{\{-x_{2}\}}-1+x_{2})\biggr] \nonumber\\
&&+\frac{2N_{2}}{N_{1}+2N_{2}}\biggl[\frac{2N_{2}}{x_{2}^2}(P_{0}-1)\exp{\{-2x_{1}\}}
(\exp{\{-x_{2}\}}-1)^2\biggr] \nonumber\\ &&+\frac{2N_{2}}{N_{1}+2N_{2}}
\frac{2N_{2}}{x_{2}^2}
\biggl[1+(P_{0}-1)\exp{\{-x_{1}\}}\biggr]
(\exp{\{-x_{1}\}}-1)(\exp{\{-x_{2}\}}-1),
\end{eqnarray} where
\begin{equation}
S_{1}(\textbf{q})=\frac{(P_{1}-1)N_{1}}{x_{1}^2}(\exp{\{-x_{1}\}}-1)^2
+\frac{2N_{1}}{x_{1}^2}(\exp{\{-x_{1}\}}-1+x_{1}),
\label{gstar}
\end{equation}
is the structure factor of a star of $P_{1}$ arms and $N_{1}$
monomers~\cite{higgins}. This expression contains two terms. The first term of
the right-hand side expresses cross-correlations between the different arms of the star.
The second term refers to the normal Debye function for a linear polymer of
$N_1$ monomers, as found by taking the appropriate limit $P_{1}=1$. Note that the general
procedure explained above can be easily generalized to calculate the structure factors
of general branched structures like the ones studied in this paper.

\vskip 3pt
%%%%%%%%%%%%%%%%%%%%%%%%%%%%%%%%%%%%%%%%%%%
\noindent {\bf Appendix B: Summary for the scaling regimes}
\vskip 1pt In this appendix we summarize the results for the scaling regimes
of the branched and hyperbranched structures studied in this work and write
them using a simpler notation. (a), (b), (c), and (d) show the results for
Gaussian polydisperse stars, SAW polydisperse stars, Gaussian dendrimers and
SAW dendrimers, respectively.

First intermediate regime $S(q)\sim q^{-s} $:

(a) $s=4(1- {1\over 2m})$

(b) $s={10\over 3}(1 - {2\over 6m-1})$

(c) $s=4(1+{1\over 2m})$

(d) $s={10\over 3}(1 + {2\over 6m+1})$

Transition points between the Guinier and the first intermediate regime:

(a) $(qR_g)^{2}\sim 1$

(b) $(qR_g)^{5/3}\sim 1$

(c) $(qR_g)^{2}\sim 1$

(d) $(qR_g)^{5/3}\sim 1$

Transition points between the first and second intermediate regimes:

(a) $(qR_g)^{2}\sim N_{max}/N_{min}$

(b) $(qR_g)^{5/3}\sim N_{max}/N_{min}$

(c) $(qR_g)^{2}\sim {\cal N}_{max}/{\cal N}_{min}$

(d) $(qR_g)^{5/3}\sim {{\cal N}_{max}P_G^{1/3}}/{{\cal N}_{min}P^{1/3}}$

Transition points between the second intermediate regime and the region
where fluctuations dominate:

(a) $(qR_g)^{2}\sim P(N_{max}/N_{min})^{1/m}$

(b) $(qR_g)^{5/3}\sim P^{2/3}(N_{max}/N_{min})^{4\over {6m-1}}$

(c) $(qR_g)^{2}\sim P_G({\cal N}_{max}/{\cal N}_{min})^{-1/m}$

(d) $(qR_g)^{5/3}\sim P_G^{2/3}({{\cal N}_{max}P_G^{1/3}}/{{\cal
N}_{min}P^{1/3}})^{-4 \over (6m+1)}$

Transition points between the first intermediate regime and the region where
fluctuations dominate:

(a) $(qR_g)^2 \sim P^{m/(m-1)}$

(b) $(qR_g)^{5/3} \sim P^{(2/3)(6m-1)/(6m-5)}$

(c) $(qR_g)^2 \sim P_G^{m/(m+1)}$

(d) $(qR_g)^{5/3} \sim P_G^{(2/3)(6m+1)/(6m+5)}$

%%%%%%%%%%%%%%%%%%%%%%%%%%%%%%%%%%%%%%%%%%%
% now the references. delete or change fake bibitem. delete next three
%   lines and directly read in your .bbl file if you use bibtex.

% that you will use with \ref{} command in the braces of the \label{}
% command.
%

\begin{figure}
\caption{Schematic representation of (a) Polydisperse stars and (b)
Dendrimeric structures formed by  connecting linear polymers.}
\label{structures}
\end{figure}

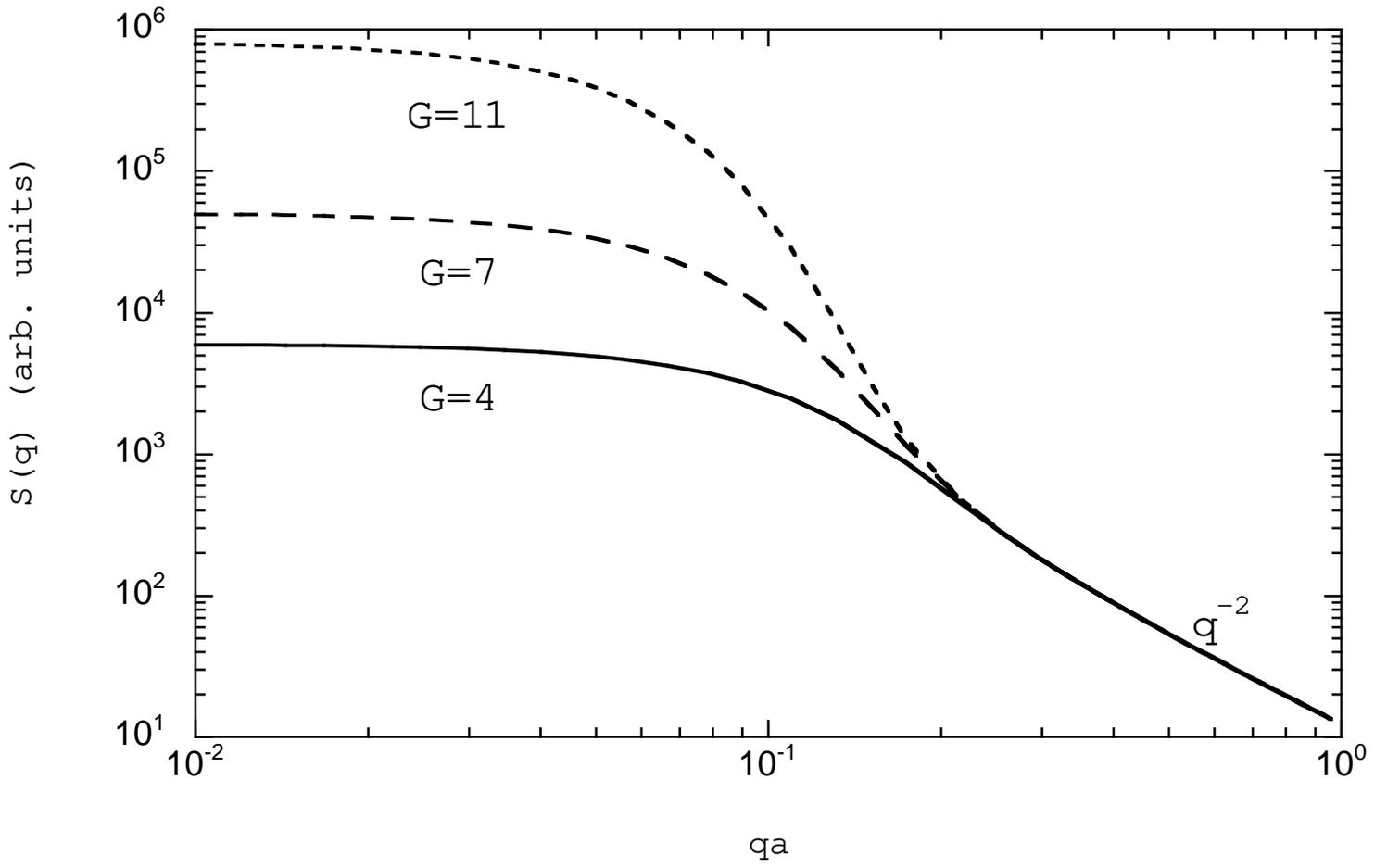
\begin{figure}
\caption{A comparison of the scattering intensity as a function of the
scattering vector in a log-log scale for a dendrimeric structure of
different number of generations, $G$. The parameters for the dendrimer are
$N_{i}=100$, $P=20$.}
\label{gdendrimer-generations}
\end{figure}

% that you will use with \ref{} command in the braces of the \label{}
% command.
% \begin{table}
% \caption{}
% \label{}
% \end{tabular}
% \end{table}


\begin{references}

\bibitem{guar} See for example, M.R. Gittings, Luca Cipelletti, V.  Trappe,
D.A. Weitz, M. In, and C. Marques, J. Phys. Chem. B {\bf 104}, 4381 (2000).

\bibitem{DLA} T.A. Witten, and L.M. Sander, Phys. Rev. Lett. {\bf 47}, 1400
(1981).

\bibitem{oh-sorensen} However see, C. Oh, and C.M. Sorensen, Phys. Rev. E
{\bf 57}, 784 (1998).

\bibitem{degennes} P.G. de Gennes, and H. Hervet, J. Phys. Lett. {\bf 44},
L351 (1983).

\bibitem{murat} M. Murat, and G.S. Grest, Macromolecules {\bf 29}, 1278
(1996).

\bibitem{mendes} E. Mendes, P. Lutz, J. Bastide, and F. BouŽ, Macromolecules
{\bf 28}, 174 (1995).

\bibitem{al-muallem 1} H.A. Al-Muallem, and D.M. Knauss, J. Polym. Sci. Part
A Polym. Chem. {\bf 39}, 3547 (2001).

\bibitem{knauss} D.M. Knauss, H.A. Al-Muallem, T. Huang, and D.T. Wu,
Macromolecules   {\bf 33}, 3557 (2000); D.M. Knauss, and H.A. Al-Muallem, J.
Polym. Sci. Part A Polym. Chem.  {\bf 38}, 4289 (2000); H.A. Al-Muallem, and
D.M. Knauss, J. Polym. Sci. Part A Polym. Chem. {\bf 39}, 152 (2001).

\bibitem{sunder} A. Sunder, J. Heinemann, and H. Frey, Chem. Eur. J {\bf 6},
2499 (2000).

\bibitem{muchtar} Z. Muchtar, M. Schappacher, and A. Deffieux,
Macromolecules {\bf 34}, 7595 (2001).

\bibitem{star} C.M. Marques, D. Izzo, T. Charitat, and E. Mendes, Eur. Phys.
J. B {\bf 3}, 353 (1998).

\bibitem{abramowitz} M. Abramowitz, and I.A. Stegun, {\em Handbook of
Mathematical Functions} (Dover Publications Inc., New York, 1972), p 297.

\bibitem{daoud-cotton} M. Daoud, and J.P. Cotton, J. Phys. France {\bf 43},
531 (1982).

\bibitem{plate} Asymptotically, for very high arm number, the high
$q$-regime of slope
$-5/3$ does not cross over directly into the intermediate regime of slope
$-10/3$ at
$q R_{\rm star}= P^{2/5}$. In fact, the semidilute character of the corona
shows up as a plateau for wavevectors smaller than the reciprocal of an
average blob size
$\xi\sim R_{\rm star} P^{-1/2}$. The scattering intensity should then be a
constant in the range
$P^{9/20}\ll q R_{\rm star} \ll P^{1/2}$. However, the ratio of the upper to
the lower boundaries is only of the order of $P^{1/20}$, and therefore
invisible under normal conditions.

\bibitem{highp} We study the limit of high number of arms and large number
of arm-sizes for which asymptotic shapes are well developed.

\bibitem{concentration} Strictly speaking, Eq.(\ref{gstardaoud1}) only leads
to a scaling form if the integral there converges. This can be achieved by
using a soft cutoff as for instance $\phi (r)=\exp(-r/r_{max})/r^{\mu}$. In
this case $S(q) \sim q^{2(\mu -3)} (q r_{max})^{2(2- \mu)} (1+(q
r_{max})^{2})^{\mu -2} \Gamma^{2}(2-\mu) \sin^{2}((2-\mu) \arctan (q
r_{max}))$ for $\mu < 3$. If
$r_{max} \gg 1$ then $S(q) \sim q^{2(\mu-3)}$.

\bibitem{prosa} T.J. Prosa, B.J. Bauer, and E.J. Amis, Macromolecules {\bf
34}, 4897 (2001).

\bibitem{higgins} J.S. Higgins and H.C. Beno\^\i t,
{\em Polymers and neutron scattering}, (Oxford University Press, Oxford,
1996).

\end{references}
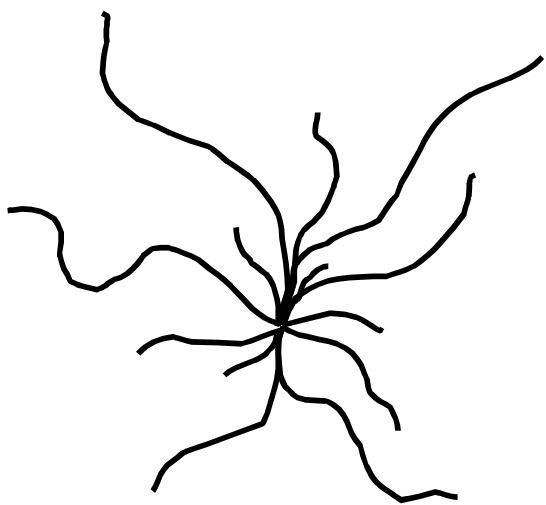
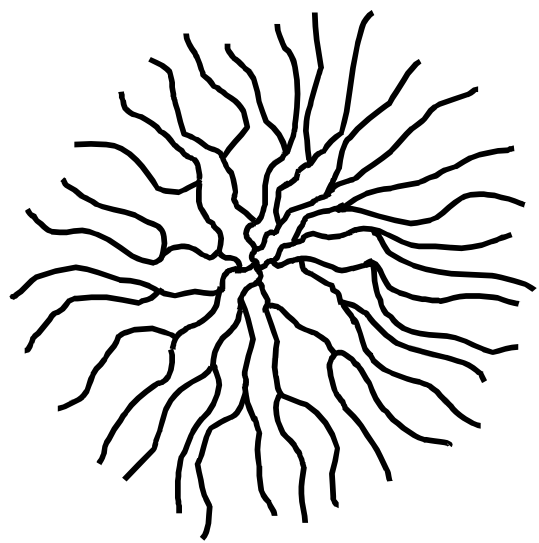
\end{document}